# Photonic crystal microdisk lasers with exceptional spontaneous emission coupling to the lasing mode

Yi-Kuei Wu[1], Xin Tu[1,2], Yi-Hao Chen[1], and L. Jay Guo[1*]

[1] *Center for Nanophotonics and Spintronics, Department of Electrical Engineering and Computer Science, The University of Michigan, Ann Arbor, Michigan 48109*

[2] *Key Laboratory for Micro and Nano Photonic Structures (Ministry of Education),Department of Optical Science and Engineering, Fudan University, Shanghai 200433, P. R. China*

*\* guo@umich.edu*

**Abstract**: A long existing issue for microdisk lasers is the small spontaneous emission coupling rate (β) in microdisk,. We found that it is due to the coupling of emitter spontaneous emission into competing Fabry-Perot (FP) modes in the microdisk cavity. Based on this analysis, we propose a new type of photonic crystal microdisk (PCM) laser to drastically suppress the photonic density of states in the vertical FP modes. As a result, we obtained a three fold increase of β than previously reported highest β of 0.15 at room temperature. The spontaneous emission control in truncated photonic crystals becomes efficient when the period number is more than one. We also demonstrate a three-stack PCM laser at low temperature with β as high as 0.72, which is 24 times higher than that in typical single microdisk (0.03), which also results in 50% threshold reduction.

**References and links**

1. S. Ishii, A. Nakagawa, and T. Baba, "Modal characteristics and bistability in twin microdisk photonic molecule lasers," IEEE Journal of Selected Topics in Quantum Electronics **12,** 71-77 (2006).
2. L. Liu, R. Kumar, K. Huybrechts, T. Spuesens, G. Roelkens, E. J. Geluk, T. de Vries, P. Regreny, D. Van Thourhout, R. Baets, and G. Morthier, "An ultra-small, low-power, all-optical flip-flop memory on a silicon chip," Nature Photonics **4,** 182-187 (2010).
3. R. Won and M. Paniccia, "Integrating silicon photonics," Nature Photonics **4,** 498-499 (2010).
4. D. J. Blumenthal, J. Barton, N. Beheshti, J. E. Bowers, E. Burmeister, L. A. Coldren, M. Dummer, G. Epps, A. Fang, Y. Ganjali, J. Garcia, B. Koch, V. Lal, E. Lively, J. Mack, M. Mašanović, N. McKeown, K. Nguyen, S. C. Nicholes, H. Park, B. Stamenic, A. Tauke-Pedretti, H. Poulsen, and M. Sysak, "Integrated Photonics for Low-Power Packet Networking," IEEE J. Sel. Top. Quant. **17,** 458 (2011).
5. F. N. Xia, L. Sekaric, and Y. Vlasov, "Ultracompact optical buffers on a silicon chip," Nature Photonics **1,** 65-71 (2007).
6. M. T. Hill, H. J. S. Dorren, T. de Vries, X. J. M. Leijtens, J. H. den Besten, B. Smalbrugge, Y. S. Oei, H. Binsma, G. D. Khoe, and M. K. Smit, "A fast low-power optical memory based on coupled micro-ring lasers," Nature **432,** 206-209 (2004).
7. J. Bleuse, J. Claudon, M. Creasey, N. S. Malik, J. M. Gerard, I. Maksymov, J. P. Hugonin, and P. Lalanne, "Inhibition,

Enhancement, and Control of Spontaneous Emission in Photonic Nanowires," Phys. Rev. Lett. **106,** 103601 (2011).

8. R. Chen, T. T. D. Tran, K. W. Ng, W. S. Ko, L. C. Chuang, F. G. Sedgwick, and C. Chang-Hasnain, "Nanolasers grown on silicon," Nature Photonics **5,** 170-175 (2011).

9. R. M. Ma, R. F. Oulton, V. J. Sorger, G. Bartal, and X. A. Zhang, "Room-temperature sub-diffraction-limited plasmon laser by total internal reflection," Nature Materials **10,** 110-113 (2011).

10. M. T. Hill, M. Marell, E. S. P. Leong, B. Smalbrugge, Y. C. Zhu, M. H. Sun, P. J. van Veldhoven, E. J. Geluk, F. Karouta, Y. S. Oei, R. Notzel, C. Z. Ning, and M. K. Smit, "Lasing in metal-insulator-metal sub-wavelength plasmonic waveguides," Optics Express **17,** 11107-11112 (2009).

11. Y. Akahane, T. Asano, B. S. Song, and S. Noda, "High-Q photonic nanocavity in a two-dimensional photonic crystal," Nature **425,** 944-947 (2003).

12. S. Reitzenstein, T. Heindel, C. Kistner, A. Rahimi-Iman, C. Schneider, S. Hofling, and A. Forchel, "Low threshold electrically pumped quantum dot-micropillar lasers," Appl. Phys. Lett. **93,** 061104 (2008).

13. K. Srinivasan, M. Borselli, and O. Painter, "Cavity Q, mode volume, and lasing threshold in small diameter AlGaAs microdisks with embedded quantum dots," Optics Express **14,** 1094-1105 (2006).

14. Q. Song, H. Cao, S. T. Ho, and G. S. Solomon, "Near-IR subwavelength microdisk lasers," Appl. Phys. Lett. **94,** 061109 (2009).

15. K. J. Vahala, "Optical microcavities," Nature **424,** 839-846 (2003).

16. J. Van Campenhout, P. Rojo-Romeo, P. Regreny, C. Seassal, D. Van Thourhout, S. Verstuyft, L. Di Cioccio, J. M. Fedeli, C. Lagahe, and R. Baets, "Electrically pumped InP-based microdisk lasers integrated with a nanophotonic silicon-on-insulator waveguide circuit," Optics Express **15,** 6744-6749 (2007).

17. J. Vuckovic, O. Painter, Y. Xu, A. Yariv, and A. Scherer, "Finite-difference time-domain calculation of the spontaneous emission coupling factor in optical microcavities," IEEE J. Quant. Electron. **35,** 1168-1175 (1999).

18. Y. Xu, R. K. Lee, and A. Yariv, "Finite-difference time-domain analysis of spontaneous emission in a microdisk cavity," Physical Review A **61,** 033808 (2000).

19. J. P. Zhang, D. Y. Chu, S. L. Wu, S. T. Ho, W. G. Bi, C. W. Tu, and R. C. Tiberio, "Photonic-wire laser," Phys. Rev. Lett. **75,** 2678-2681 (1995).

20. P. Jaffrennou, J. Claudon, M. Bazin, N. Malik, S. Reitzenstein, L. Worschech, M. Kamp, A. Forchel, and J. Gerard, "Whispering gallery mode lasing in high quality GaAs/AlAs pillar microcavities," Appl. Phys. Lett. **96,** 071103 (2010).

21. Y. Zhang, C. Hamsen, J. T. Choy, Y. Huang, J. Ryou, R. D. Dupuis, and M. Loncar, "Photonic crystal disk lasers," Opt. Lett. **36,** 2704-2706 (2011).

22. S. Noda, "Seeking the ultimate nanolaser," Science **314,** 260-261 (2006).

23. M. T. Hill, Y. S. Oei, B. Smalbrugge, Y. Zhu, T. De Vries, P. J. Van Veldhoven, F. W. M. Van Otten, T. J. Eijkemans, J. P. Turkiewicz, H. De Waardt, E. J. Geluk, S. H. Kwon, Y. H. Lee, R. Notzel, and M. K. Smit, "Lasing in metallic- Coated nanocavities," Nature Photonics **1,** 589-594 (2007).

24. J. Claudon, J. Bleuse, N. S. Malik, M. Bazin, P. Jaffrennou, N. Gregersen, C. Sauvan, P. Lalanne, and J. M. Gerard, "A highly efficient single-photon source based on a quantum dot in a photonic nanowire," Nature Photonics **4,** 174-177 (2010).

25. L. A. Coldren and S. W. Corzine, "Diode lasers and Photonic Integrated Circuits," in , Anonymous (John Wiley & Sons, 1995).

26. W. Fang, J. Y. Xu, A. Yamilov, H. Cao, Y. Ma, S. T. Ho, and G. S. Solomon, "Large enhancement of spontaneous emission

rates of InAs quantum dots in GaAs microdisks," Opt. Lett. **27,** 948-950 (2002).

27. M. D. Barnes, W. B. Whitten, S. Arnold, and J. M. Ramsey, "Homogeneous Linewidths of Rhodamine-6g at Room-Temperature from Cavity-Enhanced Spontaneous Emission Rates," J. Chem. Phys. **97,** 7842-7845 (1992).
28. H. Yokoyama and K. Ujihara, "Spontaneous emission and laser oscillation in microcavities," in , Anonymous (CRC Press, 1995).
29. S. Noda, M. Fujita, and T. Asano, "Spontaneous-emission control by photonic crystals and nanocavities," Nature Photonics **1,** 449-458 (2007).
30. Y. Zhang, M. Khan, Y. Huang, J. Ryou, P. Deotare, R. Dupuis, and M. Loncar, "Photonic crystal nanobeam lasers," Appl. Phys. Lett. **97,** 051104 (2010).
31. D. Englund, D. Fattal, E. Waks, G. Solomon, B. Zhang, T. Nakaoka, Y. Arakawa, Y. Yamamoto, and J. Vuckovic, "Controlling the spontaneous emission rate of single quantum dots in a two-dimensional photonic crystal," Phys. Rev. Lett. **95,** 013904 (2005).
32. J. T. Robinson, C. Manolatou, L. Chen, and M. Lipson, "Ultrasmall mode volumes in dielectric optical microcavities," Phys. Rev. Lett. **95,** 143901 (2005).
33. J. M. Bendickson, J. P. Dowling, and M. Scalora, "Analytic expressions for the electromagnetic mode density in finite, one-dimensional, photonic band-gap structures," Physical Review E **53,** 4107-4121 (1996).
34. Z. N. Wang, T. R. Zhai, L. Lin, and D. H. Liu, "Effect of surface truncation on mode density in photonic crystals," Journal of the Optical Society of America B-Optical Physics **24,** 2416-2420 (2007).
35. T. Baba and D. Sano, "Low-threshold lasing and purcell effect in microdisk lasers at room temperature," IEEE Journal of Selected Topics in Quantum Electronics **9,** 1340-1346 (2003).

1. **Introduction and motivation**

Laser is an essential component in the optical communication, information storage [1] [2], and optical interconnect [3]. In these research fields, in order to achieve densely integrated photonic circuits, researchers have been pursuing semiconductor lasers in miniature dimension [4][5]. Consequently, semiconductor nanocavity laser has gained considerable attention [1] [2] [6]. Four categories of semiconductor nanocavity laser have been investigated in the past years, which are based on nanowires [7][8], plasmonics [9] [10], photonic crystals [11] [12], and microdisks [13] [14].

Ideally, an ultrahigh-efficiency semiconductor laser should have the following characteristics – high quality factor (Q), large spontaneous emission coupling factor (β), small device volume, and ease of integration. β is defined as the ratio of spontaneous emission into lasing mode to the total radiative emission; a large β will lead to reduced threshold for lasing. In this regard, different types of cavities mentioned above each has its own advantages and limitations. Microdisk laser is one of the promising nanocavity lasers [15], suitable for circuit integration [2] [16], with low cavity loss [13] and compact device size [14]. However current microdisks have unexceptional spontaneous emission coupling rate (β) [17] [18], e.g. the recent reported β for the microdisk laser was less than 0.15 [19] [20] [21]. The low β impedes the realization of high-performance light-emitting devices, such as thresholdless laser [22] and ultrahigh-efficiency single photon emitters [23] [24]. Moreover, there have been little progress and lack of clear direction as to how to improve β in microdisks [17] [18], especially in experiments [19] [21].

In this letter, based on the microdisk cavity, we propose a new cavity structure to largely improve spontaneous emission coupling efficiency to the lasing whispering gallery mode (WGM) mode in a microdisk. The vertical structure is a stack in the form of truncated one dimensional photonic crystal. Such laser structure will be termed Photonic Crystal Microdisk laser, or PCM laser in short. Room temperature lasing under optical pump has been achieved in the new PCM laser even with a diameter as small as 1.22 μm, and with a small mode volume of 0.06μm$^3$. Such nanocavity laser, operated at room temperature, has a record high β of 0.5. Moreover, we show that more than one period of photonic crystal leads to strong spontaneous emission manipulation; and β is proportional to the number of the period. β for the three stack PCM laser can achieve up to a remarkable value of 0.72. Additionally, we show that the high β can decrease lasing threshold by more than 50% on InAs quantum dot laser device as compared with the single microdisk cavity, whose β is 24 times smaller than a three stack PCM.

To analyze the origin of the small β for current microdisks and how to improve it, we start with rate equations that govern the photon and carrier dynamics in a laser cavity [23][25]:

$$\frac{dN}{dt} = \eta \frac{I}{h\nu} - AN - \gamma_{lasing}BN^2 - \gamma_{nonlasing}BN^2 - CN^3 - \Gamma GS \tag{1}$$

$$\frac{dS}{dt} = \Gamma GS + \gamma_{lasing}BN^2 - \frac{1}{\tau_{ph}}S \tag{2}$$

where N is carrier density, S photon density, $\eta$ excitation efficiency; A,B and C are the coefficients for surface recombination, spontaneous emission, and Auger recombination, respectively; $\Gamma G$ represents the modal gain, and $\tau_{ph}$ the photon lifetime. Spontaneous emission coupling rate β is represented in the following equation:

$$\beta = \frac{\gamma_{lasing}}{\gamma_{lasing}+\gamma_{nonlasing}} \tag{3}$$

, where $\gamma_{lasing}$ and $\gamma_{nonlasing}$ are spontaneous emission enhancement factor into lasing and nonlasing modes, respectively [23][26][27]. They are expressed as follows:

$$\gamma = \frac{\Gamma}{\Gamma_0} = \int F_{ave}\xi(\lambda)\frac{1}{1+4Q^2(\lambda/\lambda_{cav}-1)^2}d\lambda, \tag{4}$$

where $\xi(\lambda)$ is normalized gain spectrum, Q the quality factor, and $F_{ave}$ the average Purcell factor, defined as average enhanced radiative rate of all emitters within a cavity relative to its value in free-space, which is defined differently from $F_{cav}$, maximal Purcell factor (detail in Part 1 of supplementary material). Writing this term explicitly is important in analyzing microcavity lasers because β is related to $\gamma$. According to Eq. (3), there are two extreme cases of the β [28]: $\gamma_{lasing} \gg \gamma_{nonlasing}$ and $\gamma_{lasing} \ll \gamma_{nonlasing}$. The former shows dominant spontaneous emission into lasing mode over nonlasing modes, whereas the latter vice versa.

However, microdisks do not belong to either of the above cases. In microdisk laser, $\gamma_{lasing}$ and $\gamma_{nonlasing}$ are comparable, which lead to strong competition on spontaneous emission coupling between lasing mode and nonlasing modes. For instance, $F_{ave}$ and cavity mode Purcell factor $F_{cav}$ for the WGM TE (10,1) mode in a single microdisk with diameter of 1.4um is calculated. Here 10 and 1 denote the azimuthal and radial numbers, respectively. This WGM mode has a resonant wavelength of 1.09μm and passive Q of 8600. The $F_{cav}$ of this WGM mode is 171, while $F_{ave}$ is only 1.7. This close-to-unity $F_{ave}$ leads to small $\gamma_{lasing}$ of WGM modes. On the other hand, in a microdisk cavity, apart from the lateral WGM modes with the dielectric confinement in the vertical direction, there are also cavity modes localized at the center of the microdisk that are vertically confined by the Fabry-Perot (FP) resonances such as HE10 and HE11 having similar $F_{ave}$ to that of the TE (10,1) WGM (detail in Part 2 of supplementary material). The $F_{cav}$ and $F_{ave}$ of HE10 mode are 20 and 2, while those of HE11 are 14 and 1.82, respectively. If WGM is the desired lasing modes, these FP modes with comparable $F_{ave}$ become competing modes, and therefore will significantly reduce the ratio of the spontaneous emission to the lasing modes. It is worth noting that the clarification of $F_{cav}$ and $F_{ave}$ helps explain huge discrepancy between the calculated and experimental β in [20]. Since the spontaneous enhancement factor $\gamma_{lasing}$ over the whole disk is significantly moderated (or diluted) by the spatial mismatch between emitters and WGM mode distribution, β should be estimated based on $F_{ave}$, but not $F_{cav}$. Hence the overestimation of the enhancement factor leads to inaccuracy in estimating the β.

The primary reason for small β of WGM in microdisks is stated in the previous paragraph: strong competition between lasing and nonlasing modes. Clearly, to significantly improve β of the lasing WGM mode, suppressing radiation into all the non-desirable modes becomes crucial. This can be done by reducing $\gamma_{nonlasing}$. According to Fermi's golden rule [28], manipulation of the photonic density of states controls spontaneous emission. Therefore, large β to the lasing mode can be achieved by minimizing the mode density of the other competing and non-lasing modes, e.g., the FP modes in a microdisk cavity.

Photonic Crystal (PhC) structures are well known for their abilities to modify the photonic density of states (DOS) by engineering the photonic band structures. The mode density is strongly suppressed in the wavelength region within the forbidden band. It has been demonstrated that PhCs can manipulate optical mode density and the spatial distribution of the optical modes relative to the emitter by varying the position of the photonic bandgap [29]. In particular, photonic bandgap provides a promising approach to suppress the non-lasing modes efficiently [29] [30] [31] [32]. By inhibiting spontaneous emission into nonlasing modes, the spontaneous emission rate into the lasing mode will be enhanced. It has been reported to achieve very high β in one dimensional photonic crystal (1D PhC) laser [30].

## 2. Mode density suppression in truncated one dimensional photonic crystals

Theoretically, the mode density can be modified effectively even by truncated 1D PhC [33], where the PhC structure is terminated by a bulk material with only finite pairs. Herein we first quantitatively investigate the mode density modification in a 1D PhC with different high-index and low-index pairs. Based on COMSOL simulation and calculation in [34], the mode density with various numbers of the 1D PhC pair shows in Fig. 1(a). The parameters used in this calculation are period 600nm and duty cycle 50%. As expected, the DOS in a single disk is flat within the spectrum of interest, and therefore does not have effect on mode density suppression. This result indicates that, instead of coupling into lateral propagation modes, photons in a single stack microdisk prefer to emit through the channel of Fabry-Perot modes along vertical direction. Therefore, it leads to reducing the spontaneous emission in the WGM modes. In contrast, 1D PhC with more than two pairs of high-low index stacks can strongly suppresses the DOS along the vertical direction between the wavelength of 1050nm and 1400nm, where it is located in the photonic bandgap.

To experimentally prove this idea, we conducted a micro-photoluminescence (µPL) measurement on a device shown in Fig. 1(b). The epitaxial structure of GaAs/InGaAs QWs/GaAs /$Al_{0.75}Ga_{0.25}As$ /GaAs/ $Al_{0.75}Ga_{0.25}As$ was grown with metal-organic vapour-phase epitaxy (MOVPE) technique on an n+ GaAs substrate by Landamark Technology Company. Two quantum wells (QWs) are located at the center of the top GaAs stack. The device is fabricated with selective wet etching of the AlGaAs layer to form undercut structures. The region with undercut forms truncated PhC due to large refractive index difference between the GaAs layers and the air gaps. The sample is measured with an excitation laser (Ti:Sapphire pulse laser) at region I and region II. By using the signal from region II as reference (detail in Part 3 of supplementary material), the density of states (DOS) in region I is characterized by the ratio of the signal from region I to that from region 2. Fig. 1(c) shows the calculated and measured ratio of DOS between region I (two pairs of GaAs/air stacks) and II. Both simulation and experiment show a spectrum valley between 900nm and 990nm. This is a clear evidence of DOS suppression in the 1D PhC structure. The mismatch between simulation and the measurement results at wavelength of 975 nm might be due to dramatic refractive index change of quantum wells, which is not considered in the simulation. On the other hand, the DOS for a device with single GaAs layer is shown in Fig. 1(c), which depicts a flat spectrum DOS ratio. Clearly, the DOS can be effectively modified in the photonic crystals with more than one pairs of dielectric stacks. The suppression of DOS in the vertical direction will help spontaneous emission coupling to the WGM modes more effectively. Also, the simulation of the spontaneous emission coupling rate into lasing WGM mode shown in the Part 4 of supplementary material leads to the same conclusion as this set of experiment.

3. **Photonic crystal microdisk (PCM) laser demonstration**

To boost β of the WGM modes in microdisk cavity, we propose a photonic crystal microdisk laser shown in Fig. 2(a). This device, consisting of multiple stacks of microdisks, is designed specifically to suppress the optical DOS in vertical direction and therefore to funnel most spontaneous emission into a lasing WGM mode. The multi-stack structure contributes to enhanced laser performance in the microdisk stack– lower threshold and larger β factor.

First we performed finite difference time domain (FDTD) simulation of a two-pair photonic crystal microdisk cavity with diameter of 1.2µm and an undercut of 400nm. Herein we assume the lasing mode is TE dominant due to the dominant gain for TE mode in the quantum wells. Within the full width at half maximum (FWHM) of the InGaAs quantum well gain spectrum from 960nm to 980nm, it only supports a single WGM mode TE(9,1) at wavelength of 970nm. Fig. 2(a) shows the top view of the mode profiles of WGM TE(9,1). Another mode is located at 995nm, which is far away from the FWHM of the gain spectrum. Therefore, we only consider the TE(9,1) at 970nm within the gain spectrum.

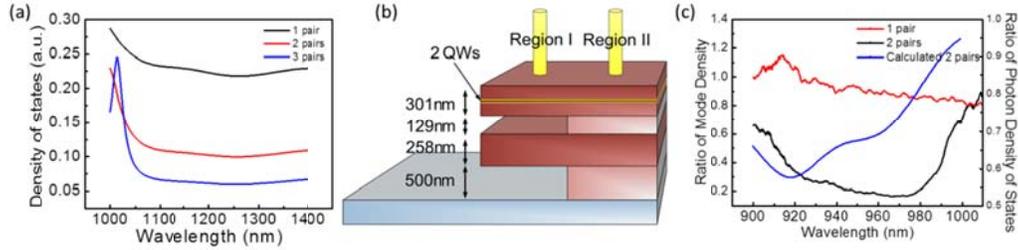

Fig. 1 (a) Mode density modification in a 1D PhC with various finite pairs (b) The schematics of the density of states measurement (c) calculated and measured ratio of DOS for single and double stack of microdisk.

We fabricated the PCM lasers (schematic shown in Fig. 2(a)) on a substrate with GaAs/AlGaAs stack structures, which has identical epitaxial structure to Fig 1(b). AlGaAs is selectively etched to make air gap as the low index layer of the PCM. A scanning electron micrograph image of the fabricated laser device is shown in Fig. 2(b). To fabricate the device, first a 300nm oxide film is deposited with plasma-enhanced chemical vapor deposition (PECVD). Then the electron beam lithography (EBL) is performed, followed by Ni deposition and a lift-off process. The resulting Nickel disk pattern is used as a hardmask in the following dry etching of the whole stack with inductively coupled plasma reactive ion etching (ICP-RIE). The PCM structure is realized with the final step of selective etching of $Al_{0.75}Ga_{0.25}As$ in dilute buffered hydrofluoric acid.

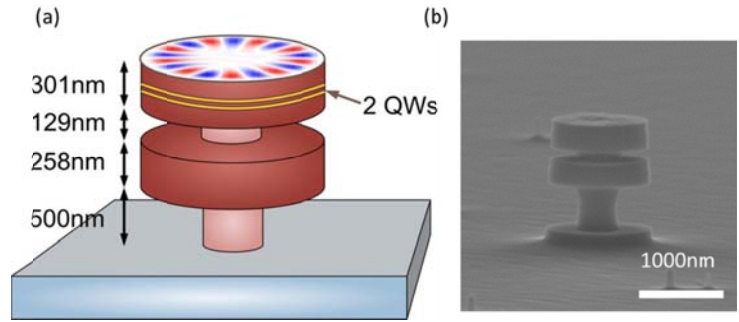

Fig. 2 (a) the schematics of the photonic crystal microdisk (PCM) laser and the field distribution of the WGM mode TE(9,1) (b) scanning electron micrograph image of PCM laser.

The measurement setup is a micro-luminescence (μ-PL) system. A Ti:Sapphire pulsed laser (Ti:Sapphire) excites the active region of the laser device through an objective lens (NA=0.9). The diameter of the excitation beam is focused down to 10μm. The signal of interest is collected by the same lens and is delivered to a monocrhometer equipped with a cooled InGaAs detector.

We demonstrated a single mode PCM laser with a disk diameter of 1.2μm at room temperature. The threshold spectrum below and above threshold is presented in the logarithmic scale in the Fig. 3(a). Single mode lasing is obtained at wavelength of 969nm, in good agreement with the simulation. The

3dB bandwidth is 1.2nm, indicating an active Q of 890. Moreover, the clear cavity peak with suppressed broad spontaneous spectrum below threshold verifies that the spontaneous emission is dominated by the lasing WGM mode in the PCM structure. Also, the bumpy background signal on the same curve indicates the suppression is not perfect in two-stack PCM due to leakage of spontaneous recombination into vertical direction. By integrating the intensity spectrum within the lasing peak, the threshold curve is obtained and illustrated in Fig. 3(b). The threshold of the two-stack PCM laser is 16mW. The purcell factor F and spontaneous emission coupling efficiency β can be fit simultaneously [35] based on the rate equations in Eq. (1) – (4), shown in Fig. 3(b). In the fitting, A=$10^4$cm/s, B=$10^{-10}$cm$^3$/s, C=$5\times10^{-29}$cm$^6$/s, and G=$2100\times v_g \times \ln(N/1.8\times10^{18})$, where $v_g$ is group velocity of light. The colored solid curves in this graph are calculated in log-log scale for purcell factor of 5.3, whereas the threshold curve in the logarithmic scale is depicted with triangular marks in this figure. β of the two-stack PML is fitted to 0.5. This is an impressive value, which is enhanced more than 3 folds than the typical value 0.15 for a single microdisk cavity [19]. This increasing β verifies again our hypothesis that the suppression of DOS in vertical direction helps improving the spontaneous emission coupling rate into WGM lasing mode. This experimental result matches with the simulated coupling rate in Part 4 of supplementary material.

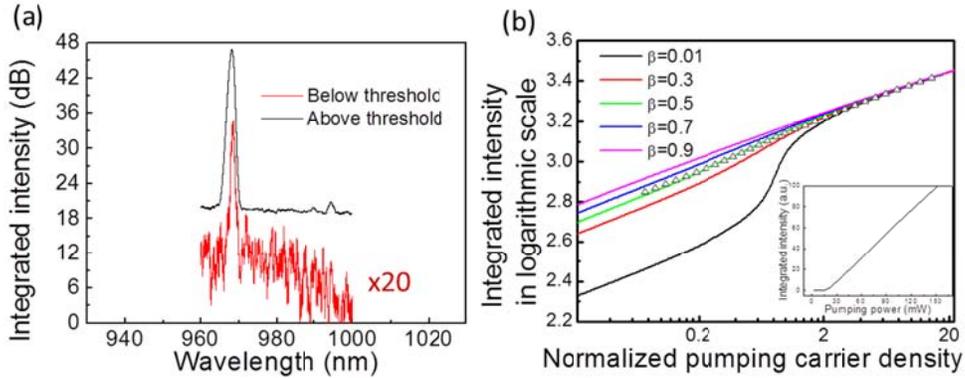

Fig. 3 (a) threshold spectrum below and above threshold of two-stack PCM in the logarithmic scale (b) The calculated threshold curve with various β(colored solid lines) and the experimental L-L curve (triangle dots). Inset illustrates the experimental L-L curve in linear scale.

4. **Discussions of PCM laser**

To further verify that multiple pairs of photonic crystal in PCM can boost β and reduce pumping threshold, we designed two laser structures, one is a three-pair photonic crystal microdisk cavity, and the other is a single microdisk. To make a fair comparison, we designed these two devices with the same purcell factor for the lasing WGM mode and same device diameter, shown in the inset of Fig. 4(a) and (b). The active Q of the lasing modes in the three-stack cavity and the single-stack cavity are 900 and 700, respectively. Both lasing modes also have comparable mode volume, 0.22 and 0.26 μm$^3$. The gain medium is self-assembled InAs quantum dots embedded in GaAs, and this measurement is performed at the low temperature (77K) and the wavelength of 1100nm. The output laser intensities are presented as a function of incident power for the three-stack and single-stack devices in Fig. 4(c) and (d), respectively. By fitting with the rate equation, it is found that the three-stack device has much larger β of 0.72 than that of the single disk device, 0.03. The small β of 0.03 is the result of more allowed vertical FP modes in competition with WGM lasing mode since the single disk is 1000 nm thick, more than three times thicker than typical reported microdisk cavity [17] [18] [19]. In this case, the DOS of the vertical FP modes is more abundant, creating even more competition channels for the spontaneous emission. It is also noticed that the three-stack PCM in this design does not achieve β near unity because of three close-spaced modes with similar quality factors have mode competition. Furthermore, the three-stack photonic crystal microdisk laser has a threshold 50% lower than that of the single layer microdisk laser. The three-stack PCM has threshold of 16mW, whereas single-stack microdisk laser has threshold of 36mW. This comparison experiment clearly demonstrates the enhancement of β and reduction of threshold by using PCM laser cavities.

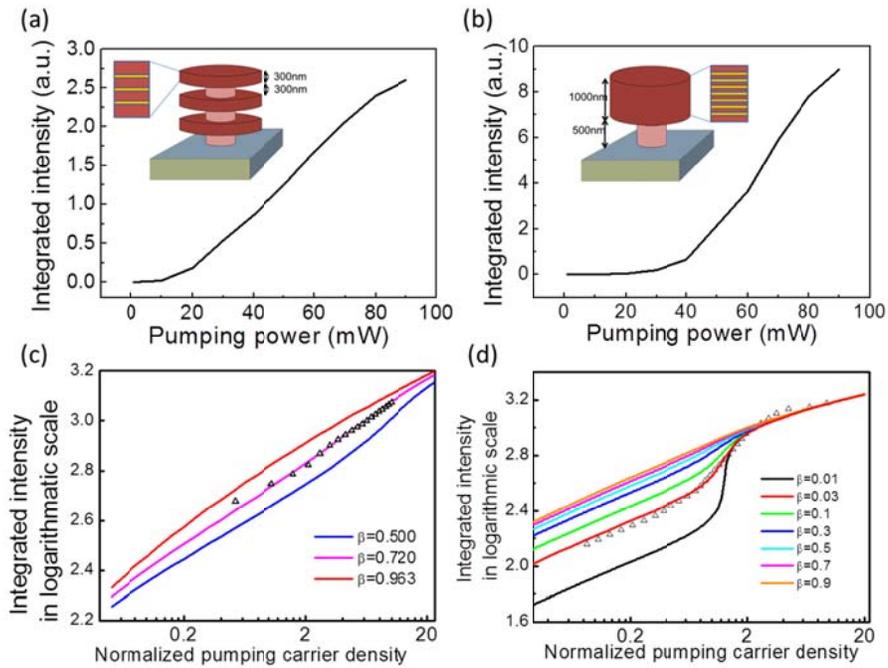

Fig. 4 The L-L curve for (a) three stack and (b) single stack microdisk lasers. The L-L curve in logarithmic scale and β fitting curves for (c) three-stack and (d) single-stack microdisk lasers. Triangular data points indicate the measured data, and solid curves are simulation result.

5. Conclusions

In conclusion, we have shown that photonic crystal microdisk laser cavity provides a large spontaneous emission coupling rate into whisepering gallery mode. This is achieved because the photonic density of states corresponding to the vertical FP resonances is effectively suppressed. We have also demonstrated that more than one pair of photonic crystal can give strong modification of optical density of states in the direction normal to the disk. With a greatly increased spontaneous coupling rate to the lasing WGMs with small mode volumes, this structure may pave the way for high-efficiency mcrodisk lasers such as thresholdless laser and single photon source.

## 6. Acknowledgments

This work is supported by DARPA grant W991NF-07-0313. We also acknowledge instrument support from electron microbeam analysis laboratory (EMAL) and Lurie nanofabrication facility (LNF) in University of Michigan, Ann Arbor.